\documentclass[preprint2]{emulateapj}

\usepackage{amssymb}
\usepackage{multirow}
\usepackage{amsmath}

\usepackage{graphicx}

\def\cm{\,{\rm cm}}

\def\ergscm2 {erg\,s$^{-1}$cm$^{-2}$}

\def\cm2 {cm$^{-2}$}

\def\aap {A\&A}
\def\apj {ApJ}

\shorttitle{ASASSN-15lh: A case for  a QN inside a WO SN remnant}
\shortauthors{Ouyed et al.}

\begin{document}
 
\title{The Superluminous (Type I) Supernova  ASASSN-15lh: \\ A case for  a Quark-Nova inside an Oxygen-type Wolf-Rayet supernova remnant}
 
\author{Rachid Ouyed\thanks{Email:rouyed@ucalgary.ca}, Denis Leahy, Luis Welbanks, Nicholas Koning}

\affil{Department of Physics \& Astronomy, University of Calgary, 2500 University Drive NW, Calgary, AB T2N 1N4, Canada}

\begin{abstract} 
We show that a Quark-Nova (QN; the explosive transition of a neutron star to a quark star) occurring
  a few days following the supernova explosion of   an Oxygen-type   Wolf-Rayet  (WO) star can
account for the intriguing features of  ASASSN-15lh, including its extreme energetics, its  double-peaked
 light-curve and the evolution  of its photospheric radius and temperature.  
 A two-component configuration of the homologously expanding WO remnant (an extended envelope and a compact core) is used 
 to harness  the  kinetic energy  ($>10^{52}$ ergs) of the QN ejecta.  The delay between the WO SN and the
 QN yields a large  ($\sim 10^4R_{\odot}$) envelope which when energized  by the QN ejecta/shock  gives the  first peak in our model. 
  As the envelope's photosphere recedes into the slowly expanding, hot and insulated, denser core (initially heated
  by the QN shock) a second hump emerges.  The spectrum  in our model should reflect  the composition of an WO SN remnant
    re-heated by a QN going off in its wake. 
 \end{abstract}


%*********************************************************************

\section{Introduction}

 Super-luminous, H-poor, supernovae (SLSNe-I) show 
 peak luminosities that are more than an order of magnitude  higher than  that of Type-Ia SNe and  that of other types of core-collapse SNe (e.g. \citealt{pastorello_2010,quimby_2011,inserra_2013,nicholl_2014}); see also  \citealt{galyam_2012}
 for a review). Suggested explanations include 
   powering by black hole  accretion (\citealt{dexter_2013}),  millisecond magnetars (\citealt{woosley_2010, kasen_2010}), 
    jet energy input  (\citealt{gilkis_2016}), 
    sub-millisecond strange quark stars (\citealt{dai_2016})),  tidal disruption of a star by a supermassive BH (\citealt{leloudas_2016}),
     pulsationnal pair-instability progenitor models (\citealt{kozyreva_2015,chatz_2015})
   as well as models involving the interaction of  SN ejecta with its surroundings (\citealt{wang_2015,moriya_2015,sorokina_2015,piro_2015,dessart_2015}).
 Following the discovery of the
  most luminous of them all, ASASSN-15lh (\citealt{dong_2016}),  the proposed models have been challenged and pushed to their limits because of the high energy and re-brightening of ASSASN-15lh (\citealt{brown_2016}; 
   e.g. discussion in \citealt{chatz_2016}). It is safe to say that the exact mechanism powering SLSNe-I 
   remains unknown, leaving  room for novel ideas to be explored.

     The QN model for SLSNe appeals to the explosive transition
      of a neutron star (NS) to a quark star (QS; the QN compact remnant) a few days
      following the SN explosion.  The extended envelope
   means that PdV losses are minimal
   when it is shocked by the QN ejecta yielding a SLSN. 
    This can be the case if the SN explosion of a single massive progenitor is followed
    a few days later by a QN  (\citealt{leahy_2008,ouyed_2009,ouyed_2012,kostka_2014a}).
    A NS in-spiralling inside a  Common Envelope in massive binaries gives a similar
     scenario and if the NS goes QN a SLSN-I results since the hydrogen has been
     lost during the binary evolution (\citealt{ouyed_2015a,ouyed_2015b,ouyed_2016}).
       We find that   ASASSN-15lh is best explained  if we consider
     a QN occurring a few days following the SN explosion of a  stripped oxygen sequence Wolf-Rayet star (i.e. an WO-type;  \citealt{barlow_1982}).
      A QN going off in a  two-component configuration WO SN remnant (SNR) which consists of a core (a dense, inner region)
       and an envelope (the extended lower density outer layers) seems to 
       account for the general properties
   of ASASSN-15lh's ligtcurve.  The time delay between the QN and the preceding SN means that the
   QN energy is deposited in a very extended WO SN envelope. When shocked by the QN ejecta,   the 
envelope  yields  the first hump  while the second hump results from the late and slower contribution
of the denser inner region (i.e. the core; initially  energized by the QN shock; see Figure \ref{fig:sequence} for illustration purposes). 
The heated core remains
insulated (effectively acting as a ``hot plate") until it is revealed by the receding photosphere of the QN-shocked envelope.
  Our model reproduces the light-curve of ASASSN-15lh and  the evolution of its photospheric radius and temperature.   
  The spectrum in our model should represents the composition of the  WO SNR re-heated by the QN shock.
  In \S 2,  we start with a brief overview of the physics of the QN and of  its ejecta followed by a description of the WO SN progenitor
  it collides with.    The method we use to 
 compute the light-curve and the properties of the photosphere is given in \S 3.
   We end with a  discussion and a conclusion  in \S 4.

\section{The QN in the wake of an WO SN remnant}

\subsection{The QN energetics}

 Assuming   that matter made of up, down and strange quarks  is more stable than
  hadronic matter (\citealt{bodmer_1971,witten_1984}),  the explosive conversion of a NS to a quark star (QS) remains
  an intriguing  possibility. Theoretical and numerical studies have shown that such
   a transition is not unlikely  (e.g. \citealt{dai_1995,cheng_1996,horvath_1988,ouyed_2002,keranen_2005,niebergal_2010, herzog_2013,pagliara_2013,furusawa_2015a,furusawa_2015b,drago_2015a}.
   These studies find three possible paths to the explosion including,  neutrino-driven explosion mechanisms  (\citealt{keranen_2005,drago_2015b}),
  photon-driven explosions  (\citealt{vogt_2004,ouyed_2005}) and quark-core-collapse driven explosion plausibly induced by 
       deleptonization instabilities (\citealt{niebergal_2010}; see   \citealt{ouyed_2013a} for a recent review).  More sophisticated simulations are required to
       better capture the micro-physics and macro-physics of the transition. Until then,  one may hope for  direct and/or indirect 
     observation of the QN to confirm the transition.    Here, as in some of our previous work, we assume that the transition is triggered when
   deconfinement of neutrons to quarks happens inside a massive NS.  This may be the case here, if for example, fall-back
   material from the preceding WO SN drives the NS's mass above the critical value for quark deconfinement. 
  When combined with  strange-quark seeding inside the deconfined core, ignition and a conversion
  ``flame"  can be fuelled (see \citealt{niebergal_2010}).   For a NS of mass $M_{\rm NS}$, 
   a QN can   release at least $(M_{\rm NS}/m_{\rm p}) E_{\rm conv.}\sim  10^{53}\ {\rm ergs}$  from the direct conversion 
   of its hadrons   to quarks with  
   up to $E_{\rm conv.}\sim 100$ MeV of energy released per hadron converted (e.g. \citealt{weber_2005}); $m_{\rm p}$ is the proton mass. Accounting for  gravitational  energy from contraction and additional 
   energy release during phase transitions within the quark phase (e.g. \citealt{vogt_2004}), the total energy can reach 
   a few times $10^{53}$ ergs.  The amount imparted as kinetic energy ($E_{\rm QN}^{\rm KE}$) to 
     the NS outer layers ($\sim 10^{-3}M_{\odot}$)  exceeds   $\sim 10^{52}$ ergs (e.g. \citealt{keranen_2005}).  
     The  QN ejecta is
     a very dense,  neutron-rich,  relativistic  material   expanding away from the explosion 
     point at a Lorentz factor $> 10$ (\citealt{jaikumar_2007,ouyedleahy_2009,kostka_2014b}).

 \begin{figure}[t!]
 ~\\~\\
   \includegraphics[scale=0.12]{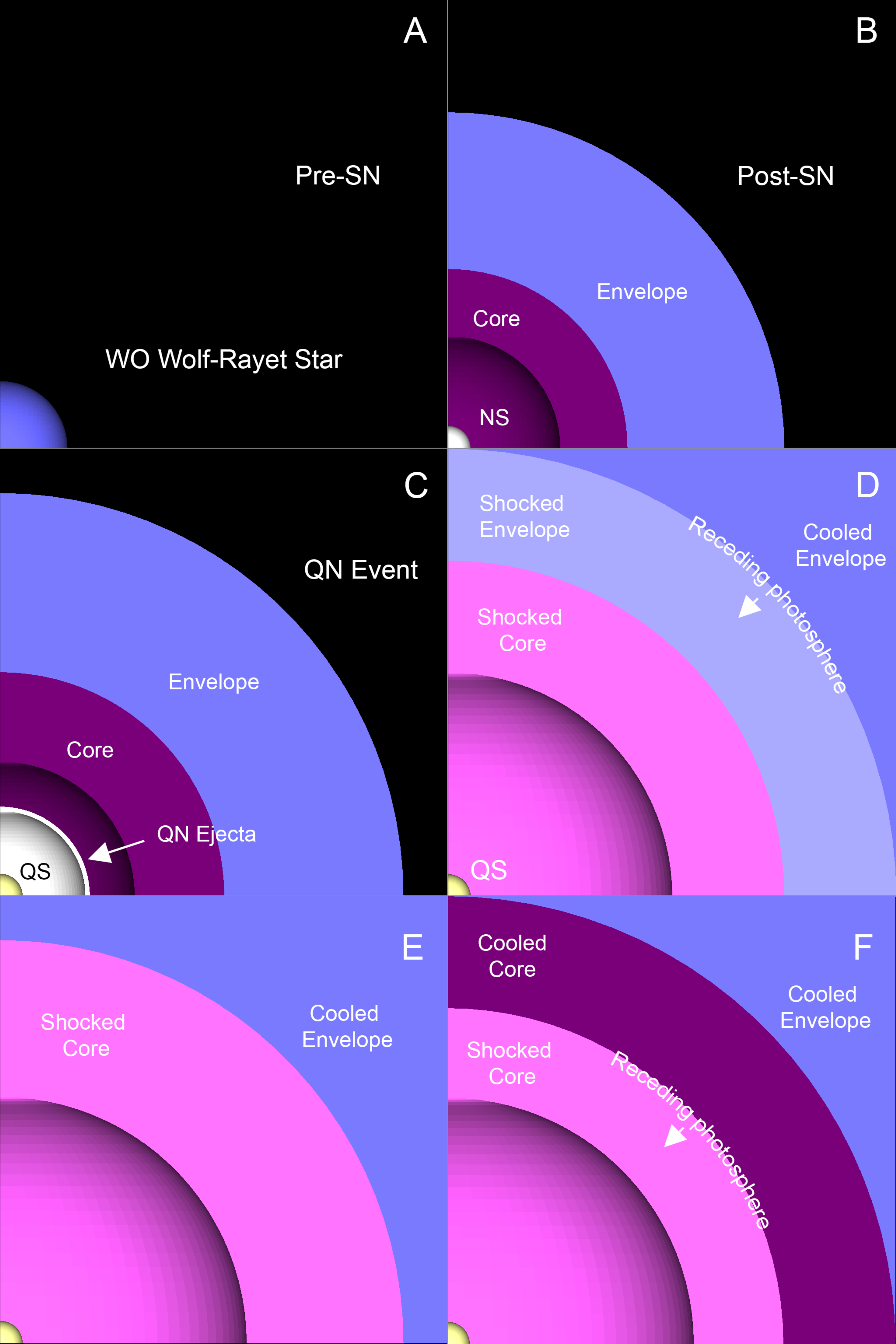}
   \caption{Sequence of events in our model starting with the SN explosion of a WO-type star (panels A \& B). Panel B depicts the homologously expanding SN ejecta a few days later (just before the QN event) which consists of a dense core and an overlaying extended, and much lower density,   envelope. The QN goes off (panel C) leaving behind a QS (the compact remnant) and generating a very dense, relativistic ejecta, which shocks the WO SN remnant. The core is shocked first and then the envelope which starts to radiate and cool as portrayed by its receding photosphere (panel D). The buried core is eventually unveiled (panel E) triggering radiation escape directly from the core giving us the second hump in our model (panel F).~\\}
 \label{fig:sequence}
 \end{figure}

\subsection{The WO supernova remnant}

The extensive mass-loss (of hydrogen and helium) experienced by Wolf-Rayet  stars unveils
 the products of He-burning ($^8$Be, $^{12}$C, $^{16}$O) at the stellar surface
  with the Oxygen-type stars showing much more oxygen than Carbon-type
  (e.g. \citealt{barlow_1982}). Consider an WO star  that just experienced a SN event (panels A \& B in Figure \ref{fig:sequence}).  If we define
$M_{\rm WO}$ as the remnant's mass and $E_{\rm SN}^{\rm KE}$ as
 the ejecta's kinetic energy, then for a homologously expanding, uniform density,  ejecta
 the expansion velocity is  $v_{\rm SN} = \sqrt{(10/3) E_{\rm SN}^{\rm KE}/M_{\rm WO}}$.
    To simplify our model, we assume that the remnant consists of a core
 of mass $M_{\rm core}$ and an envelope of mass $M_{\rm env.}= M_{\rm WO}-M_{\rm core}$.
     Defining $t_{\rm delay}$ as the time delay between the SN
  explosion and the QN explosion,  the size of the WO SN remnant when the QN goes off
  is $R_{\rm env., 0}=R_{\rm WO}+ v_{\rm SN} t_{\rm delay}\simeq v_{\rm SN} t_{\rm delay}$ 
 for a SN ejecta expanding at speed $v_{\rm SN}$ with $R_{\rm WO}$ being 
   the radius of the WO-type progenitor star which is negligible   ($R_{\rm WO} << R_{\rm env., 0}$).

     Once the QN goes off (panels C in  Figure \ref{fig:sequence}), its extremely dense and relativistic ejecta catches up very quickly with 
     the preceding SN ejecta.  For example for a typical SN expansion speed of $10^4$ km s$^{-1}$ and a time
     delay of, say, 10 days between the SN and the QN explosion the two ejecta collide after only  a fraction of an hour 
      following  the QN explosion;  i.e. $v_{\rm SN} t_{\rm delay}/c << t_{\rm delay}$ where $c$ is the speed of light.  
      In reality,  the QN ejecta hits the inner edge of the SN ejecta even much sooner since the speed
   of the inner SN ejecta is very small.

    The QN shock is  radiative  owing to 
 the low density of the envelope  meaning that most
 of the QN energy is converted to thermal (gas and radiation) and only a small portion is in the 
 form of kinetic energy. In this case,  the envelope expansion velocity after the QN shock
  is effectively that of the SN or,  $v_{\rm env.} = v_{\rm SN}$.  
  This also implies that the shock velocity in the envelope is given as $v_{\rm env., sh.}= (7/6) v_{\rm SN}$.  
  As the shock makes its way through the large WO SN remnant, part of the QN kinetic energy $E_{\rm QN}^{\rm KE}$  is 
converted to thermal energy in the core ($E_{\rm core}$)
  and the remaining  ($E_{\rm env.} =  E_{\rm QN}^{\rm KE} - E_{\rm  core}$) is used up in the envelope.
  The QN shock propagating in the envelope (we refer to it as the outer QN-shock) provides the
  first energy source as it heats up the envelope while the QN shock which propagated in the core (we refer to it as 
   the inner QN-shock) heats up the core. The radiation from these two energy sources  
    first diffuses in the envelope (panel D in  Figure \ref{fig:sequence}) and later  in the
   core (panels E \& F in  Figure \ref{fig:sequence})  yielding the double-hump nature of the light-curve in ASASSN-15lh as described in the next section.
   Finally, the SN explosion leads to oxygen- and carbon-burning products in the denser hotter layers of the WO star;
i.e. in the core. All of carbon and oxygen should be preserved in the outer layers
  (i.e. in the envelope).    A radiative QN shock is unlikely to alter   
  the chemical composition in the WO SNR which should be reflected in the spectrum
  according to our model.

 \section{The  light-curve in our model}

 \subsection{Diffusion in the envelope}
   \label{sec:qn1}

  To simulate diffusion in the envelope, we 
  integrate Eq. (3) in \citet{chatz_2012} using a `top-hat' prescription as described in  \S 3.2.1 in that paper.
  In our case,  since we have two sources contributing to the envelope the input luminosity function is given  as
    \begin{equation}
   L_{\rm env., input}  = 
\begin{cases}
    L_{\rm env., sh.} + L_{\rm core}, & \text{if }  t \le t_{\rm env., sh.}\\
    L_{\rm core},             &   \text{if } t_{\rm env., sh.}  <  t \le t_{\rm core, sh.}\\
    0,           & \text{otherwise}
\end{cases}
\end{equation}
  
  The QN shock input luminosity into the envelope is $L_{\rm env., sh.}= E_{\rm env.}/t_{\rm env., sh.}$
   for $t< t_{\rm env., sh.}$ and $L_{\rm env., sh.}=0$ otherwise.  
    The shock duration $t_{\rm env., sh.}$ 
   is the time it would take the QN shock to traverse the envelope
 and break-out. It is given as  $t_{\rm env., sh.}= (R_{\rm env., 0} + v_{\rm SN} t_{\rm env., sh.})/v_{\rm env., sh.}$.
  This gives $t_{\rm env., sh.}= 6 R_{\rm env., 0}/v_{\rm SN}\simeq 6 t_{\rm delay}$ since $R_{\rm env., 0}= R_{\rm WO} + v_{\rm SN} t_{\rm delay}\simeq v_{\rm SN} t_{\rm delay}$. 
  
  The core (first shocked by the QN ejecta) is embedded deep in the envelope and is 
    insulated radiating at a lower and slower rate 
    than the outer shock.  It effectively acts  as a ``hot plate" with an input luminosity into the envelope
    given by $L_{\rm core, sh.}= E_{\rm core}/t_{\rm core, sh.}$. 
     The time  $t_{\rm core, sh.}$ corresponds to when the envelope's photosphere 
    recedes deep enough to cross the core radius. For $t > t_{\rm core, sh.}$,
     the envelope is optically thin to the radiation from the core and diffusion switches
      into core diffusion.       The input luminosity is that from the core 
        for $t_{\rm env., sh.} < t < t_{\rm core, sh.}$ while the input luminosity
        from the outer shock dominates mainly at $t < t_{\rm env., sh.}$. 
        This is also reflected in the output luminosity. 
       The time $t_{\rm core, sh.}$ 
    is derived self-consistently  in our model as explained in \S \ref{sec:qn1} below.

  The characteristic or effective LC timescale in the envelope (not to be confused with diffusion timescale; see
  Eq. (3) in \citealt{chatz_2012})  is  $t_{\rm env., d}=  \sqrt{2 \kappa_{\rm env.} M_{\rm env.}/(\beta c v_{\rm SN})}$ with $\kappa_{\rm env.}$ the mean opacity in the envelope, $\beta=13.8$ (\citealt{arnett_1980,arnett_1982}) and $c$ the speed
  of light. The corresponding photospheric radius  is derived using $R_{\rm env., ph.}  = R_{\rm env.}(t) -  \alpha_{\lambda}  \lambda_{\rm env.} (t)= (R_{\rm env., 0}+ v_{\rm SN} t )-  \alpha_{\lambda} \lambda_{\rm env.} (t)$
where $\lambda_{\rm env.} (t) = 1/(\kappa_{\rm env.} \rho_{\rm env.}(t))$ is the photon mean-free path in the envelope whose 
 density is $\rho_{\rm env.} (t)= M_{\rm env.}/(4\pi/3~ R_{\rm env.}(t)^3)$. 
 The factor $\alpha_{\lambda}$  is meant to emulate the spherical geometry of the very extended
   SN ejecta in our model  and the complex opacity effects (temporal and chemical variability) not captured by our
   simplified model. The
 $\alpha_{\lambda}= 2/3$  value is only appropriate to planar geometry and when $\lambda_{\rm env.} << R_{\rm env.}$,
 which is  not the case here.

 \subsection{Diffusion in the core}
   \label{sec:qn1}

   The output luminosity from diffusion in the core is also found using 
   Eq. (3) in \citet{chatz_2012}  for $t> t_{\rm core, sh.}$.  The second term in that equation 
    is the  exponential decay expected when pure diffusion controls the luminosity
    output. The initial value of the output luminosity in this stage 
    is the power  output from the envelope at $t=t_{\rm core, sh.}$.  The time $t_{\rm core, sh.}$ is found when the envelope's photosphere crosses the core radius; i.e. when $R_{\rm env., ph.}= R_{\rm core}$.
     Mathematically we solve the following equation 
      $R_{\rm env., 0}+ v_{\rm SN}t -  \alpha_{\lambda} (2/3) \lambda_{\rm env.}= R_{\rm core, 0}+ v_{\rm core} t$.
       This introduces another parameter which is the initial size of the core, $R_{\rm core, 0}$ when the QN goes off.
        For a homologously expanding ejecta this gives us the core's expansion velocity from  $v_{\rm core}= v_{\rm SN} (R_{\rm core, 0}/R_{\rm env., 0})$.  The characteristic LC timescale in the core  is then  $t_{\rm core, d}=  \sqrt{2 \kappa_{\rm core} M_{\rm core}/(\beta c v_{\rm core})}$ with $\kappa_{\rm core}$ the mean opacity in the core.

    To summarize, the free parameters in our model can be divided into the following categories: (i) The SN parameters ($E_{\rm SN}$ and $M_{\rm WO}$); 
(ii)  The QN parameters ($E_{\rm QN}^{\rm KE}$ and $t_{\rm delay}$); (iii) The envelope parameters $E_{\rm env.}/E_{\rm QN}^{\rm KE}$,  
the envelope mass $M_{\rm env.}$ and the envelope's mean opacity $\kappa_{\rm env.}$; (iv) The core  initial radius $R_{\rm core, 0}/R_{\rm env., 0}$  and its mean opacity $\kappa_{\rm core}$; (v) The parameter $\alpha_{\lambda}$ related to 
 the photospheric radius.

\begin{figure}[h!]
   \includegraphics[scale=0.85]{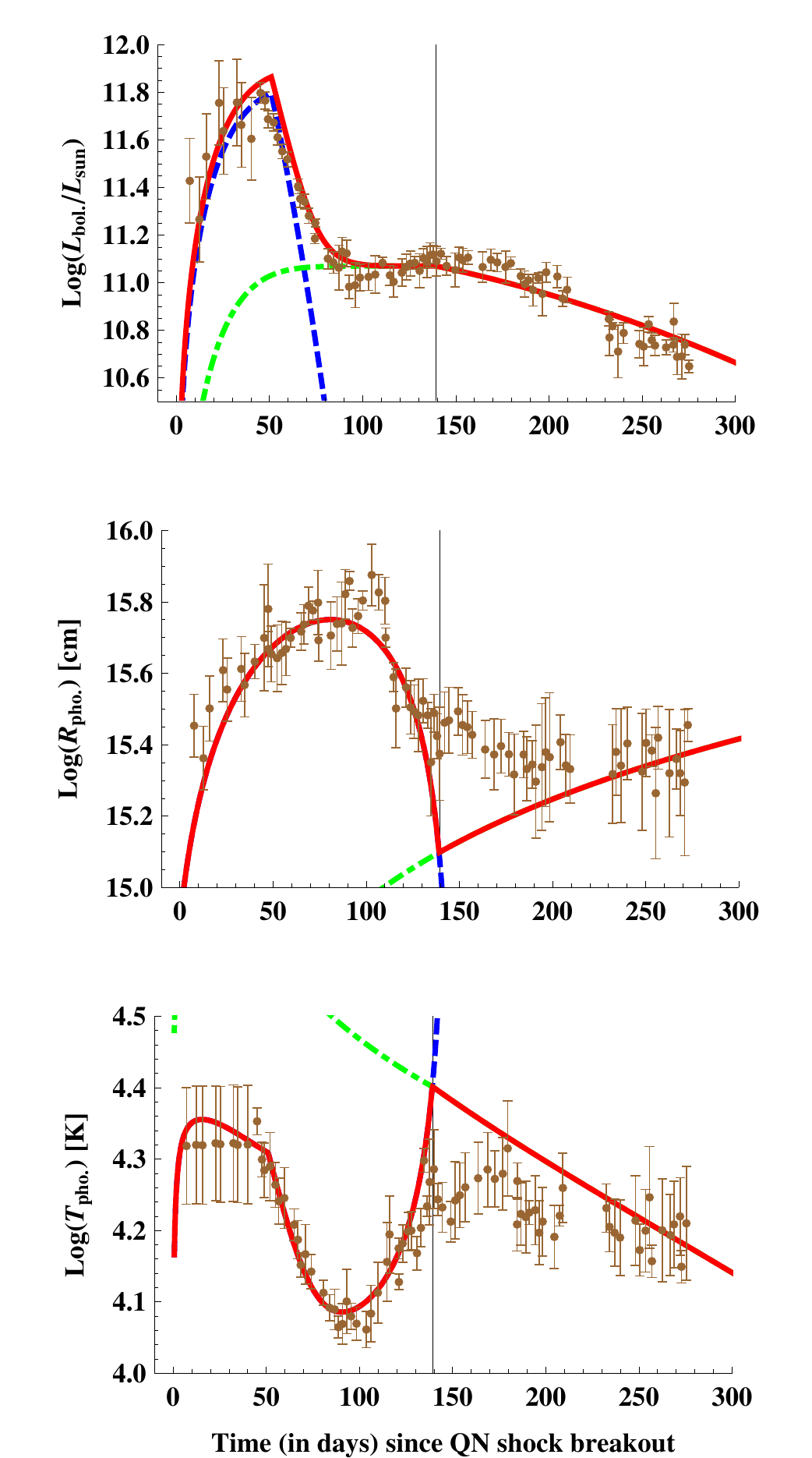}
   \caption{The QN model fit (solid red line) to ASASSN-15lh's bolometric LC (top panel),
   photospheric radius (middle panel) and photospheric temperature (bottom panel); the data is from  \citet{dong_2016}
   and \citet{brown_2016}.  The blue dashed (green dot-dashed) curve  
   depicts the envelope's (core's) contribution.
  The vertical line depicts the moment in time when the envelope's photosphere crosses
   the core. The time of explosion in our model is $30$ days relative to $t=0$ in the data.~\\}
    \label{fig:assassinfig}
 \end{figure}

 Figure \ref{fig:assassinfig} shows our fit to  ASASSN-15lh's bolometric LC, its photospheric radius and 
  temperature using the parameters given in Table \ref{table:assassintab}.
 The output luminosity ($L_{\rm out}$, shown as the solid red line in Figure \ref{fig:assassinfig}) is first dominated by that
 from the QN-shocked envelope before switching to the contribution from the  core 
 shortly after $t=t_{\rm env., sh.}\simeq 6t_{\rm delay}$. 
 When the envelope's photosphere crosses the  core at $t=t_{\rm core}$,  it becomes transparent to the core
  and the core's input power into the envelope ceases.  The luminosity
  then  is from pure diffusion in the core   declining exponentially as the core cools. 
 For the fit shown here $t_{\rm core}\simeq 140$ days which is depicted by the vertical thin line in the 
 panels  Figure \ref{fig:assassinfig}.

  Shown in the middle panel is the photosphere in the envelope (the blue dashed curve) as it reaches
  a maximum before it recedes deeper in the envelope first powered by the outer shock then by the slowly contributing
  inner shock. Eventually, the photosphere in the envelope reaches deep into the core (at $t=t_{\rm core, sh.}$) 
  at which point in time it disappears leaving place  for a new photosphere to propagate inside the core, cooling it (the green
  dot-dashed curve).   The effective photospheric radius ($R_{\rm eff.}$) is  the maximum photospheric radius between  the two phases
 (middle panel).   
  The  effective  temperature  (shown in the lower panel) is derived from the  blackbody luminosity as $T_{\rm eff.}= (L_{\rm out}/(4\pi \sigma_{\rm SB} R_{\rm eff.}^2))^{1/4}$; $\sigma_{\rm SB}$ is the Stefan-Boltzmann constant. A simultaneous fit of the LC and the photospheric radius required
     $\alpha_{\lambda}= 1.6$ and better captured the transition between the two photospheres. 
 In reality, the transition from the envelope to the core is more complex and would necessitate 
  numerical simulations to reproduce. Nevertheless, despite the simple approach we adopted here, our model is capable
  of capturing the general light-curve behavior and the corresponding effective photospheric radius and temperature.

   \begin{table*}[t!]
\begin{center}
\caption{Sample parameter fits to the  LC  of  Asassn-15lh}
 \label{table:assassintab}
%\resizebox{\columnwidth}{!}{%
\begin{tabular}{|c|c||c|c||c|c|c||c|c|}\hline
   \multicolumn{2}{|c||}{SN} &  \multicolumn{2}{|c||}{QN} &  \multicolumn{3}{|c||}{Envelope} &  \multicolumn{2}{|c|}{Core}  \\
  \hline
  $E_{\rm SN}$ (ergs)  & $M_{\rm WO}$ ($M_{\odot}$) &   $E_{\rm QN}^{\rm KE}$ (ergs) & $t_{\rm delay}$ (days) & $E_{\rm env.}/E_{\rm QN}^{\rm KE}$  & $M_{\rm env.}$ ($M_{\odot}$) & $\kappa_{\rm env.}$ (cm$^2$ g$^{-1}$)   & $R_{\rm core, 0}/R_{\rm env., 0}$    & $\kappa_{\rm core}$ (cm$^2$ g$^{-1}$)     \\\hline
 $2.5\times 10^{51}$ & 3.5 & $1.7\times 10^{52}$& 8.5 &  0.68 & 1.0 & 1.2 & 0.09 & 2.4    \\\hline
\end{tabular}
%}
~\\
 \end{center}
\end{table*}

~\\
    
\section{Discussion and conclusion}
 
  Here we put forward the idea of a QN occurring inside a homologously expanding, uniform density, SN remnant of a stripped WO star 
  to explain the observed properties of ASASSN-15lh.
   For  a time delay of $\sim 8.5$ days between the SN and the QN,  the model is capable of reproducing 
  the energetics and the  double-peaked light curve of ASASSN-15lh as well as 
    the overall behavior of the photosphere (radius and temperature).
The first  hump in our model is from the QN shock propagating in the low-density
 envelope while the second hump is from the slower contribution of the deep hot core to the envelope
 heating.  Below are some  features and predictions of our model:
 
(i) No hydrogen or helium was detected in ASASSN-15lh
(\citealt{dong_2016}) which is expected in our model  since the 
Wolf-Rayet (WO) star has been stripped of both elements prior to the SN explosion;

(ii)  If  the stripped Wolf-Rayet star experienced a standard SN shock, as we propose here, then oxygen and carbon burning should yield some magnesium, silicon, sulfur and some calcium which we predict may be present  (albeit in small quantities) in the spectrum of ASASSN-15lh-like events.
 The time delay between the SN and the QN means that the QN shock will propagate in a low-density WO remnant 
(as compared to a SN shock in a stellar envelope)
and in principle should not affect the chemical composition of the WO SNR. 
 Contribution from $^{56}$Ni decay could in principle be detected later in the LC evolution  but this may be too
 faint particularly if fall-back  during the  SN was important. The  fall-back would help the NS gain enough  mass to experience
  quark deconfinement in its core and undergo a QN explosion.
The  spectrum will be modelled and simulated  elsewhere;

 (iii) A nearby ASASSN-15lh-like event  may show a precursor, the associated WO-type SN explosion, if the time delay
   between the SN and QN is long enough  to avoid overlap with the brighter peaks.  In our model, naturally, the overall shape
of the resulting SLSN light-curve should vary from a single hump
to a triple hump  depending on the time delay between
the SN and the QN and the importance of the core's contribution.  We even
predict   a fourth hump if the QN-shocked ejecta
manages to collide with the WO wind or any circumstellar material. Such
a late bump would show signatures of hydrogen and  helium;

 (iv) While the model presented here  focuses
 on a QN in the wake of an WO-type SN, other scenarios involving QNe inside other types of WR stars (e.g. WN and WC)
 may be possible which should yield similar superluminous events with nevertheless some distinctive spectral features.
  We note that since WO stars are the rarest amongst WR stars, events such as ASASSN-15lh would be scarce;

(v)  The QN leaves behind a QS with a  spin-down (SpD) timescale  $\tau_{\rm SpD}\simeq 4\times 10^7\ {\rm s}\ P_{10}^2 B_{14}^{-2}$ if born with a 10 millisecond
 period and a typical QS magnetic field of $10^{14}$ G (e.g. \citealt{iwazaki_2005}). This gives a spin-down  X-ray luminosity $L_{\rm X}\simeq 5\times 10^{42}\ {\rm erg\ s}^{-1}  (1+ t/\tau_{\rm SpD})^{-2}$.  One year after the QN event  this would correspond to  $\sim 10^{42}$ erg s$^{-1}$ which seems to agree with the detected persistent X-ray emission in ASASSN-15lh  (\citealt{margutti_2016}). The current QS period
   exceeds  $10$ ms but spin-down power could still affect the late evolution of ASASSN-15lh LC.

  (vi)   Our model applies to any progenitor star (O-stars, luminous blue variables, Wolf-Rayet stars, etc.) if it is massive
  enough to go SN and form a heavy NS but not too massive to go directly to a black hole. This is the
  unifying scheme between SLSNe, as first suggested in \citet{ouyed_2013b} and \citet{ouyed_leah_2013}, where the Type (I or II) of the SLSN defines  the 
  chemical composition of the SN progenitor. We re-iterate, and argue again, that 
   H-poor SLSNe would  occur  in  higher-metallicity  environment  (i.e. with 
higher stellar mass loss-rates) while  low-metallicity progenitors would lose less mass and should be 
 linked to SLSN-II if a QN goes off in their wake.

%  \vskip -0.3in
\begin{acknowledgements}   
This work is funded by the Natural Sciences and Engineering Research Council of Canada. 
\end{acknowledgements}

%---------------------------------------------------------------


\begin{thebibliography}{99}
% 

   \bibitem[Arnett(1980)]{arnett_1980}     Arnett, W. D. 1980,\apj, 237, 541
    
 \bibitem[Arnett(1982)]{arnett_1982}   Arnett, W. D. 1982, \apj, 253, 785
 
  \bibitem[Barlow \& Hummer(1982)]{barlow_1982} Barlow,  M.  J.,  \&  Hummer,  D.  G.  1982,  in  Wolf-Rayet  Stars:  Observations,
Physics, Evolution, eds. C. W. H. De Loore, \& A. J. Willis, IAU Symp., 99,
387
    
    \bibitem[Bodmer(1971)]{bodmer_1971}  Bodmer, A. R. \ 1971, Phys. Rev. D, 4, 1601
    
    \bibitem[Brown et al.(2016)]{brown_2016}      Brown, P.~J., Yang, Y., Cooke, J., et al.\ 2016, \apj, 828, 3
    
   \bibitem[Chatzopoulos et al.(2012)]{chatz_2012}    Chatzopoulos, E., Wheeler, J.~C., \& Vinko, J.\ 2012, \apj, 746, 121
   
 \bibitem[Chatzopoulos et al.(2015)]{chatz_2015}      Chatzopoulos, E., van Rossum, D. R., Craig, W. J., Whalen, D. J., Smidt, J.,
\& Wiggins, B. 2015, ApJ, 799, 18, 1410.0039
   
\bibitem[Chatzopoulos et al.(2016)]{chatz_2016} Chatzopoulos, E., Wheeler, J. C., Vinko, J., Nagy, A. P., Wiggins, B. K., \&
Even, W. P. 2016, ArXiv e-prints

\bibitem[Cheng \& Dai(1996)]{cheng_1996} Cheng, K.~S., \& Dai, Z.~G.\ 1996, Physical Review Letters, 77, 1210

\bibitem[Dai et al.(1995)]{dai_1995} Dai, Z., Peng, Q.,  \& Lu, T.\ 1995, \apj, 440, 815

\bibitem[Dai et al.(2016)]{dai_2016} Dai, Z.~G., Wang, S.~Q., Wang, J.~S., Wang, L.~J., \& Yu, Y.~W.\ 2016, \apj, 817, 132 

 \bibitem[Dessart et al.(2015)]{dessart_2015}   Dessart, L., Audit, E., \& Hillier, D. J. 2015, MNRAS, 449, 4304,
1503.05463

\bibitem[Dexter \& Kasen(2013)]{dexter_2013}   Dexter, J. \& Kasen, D. 2013, \apj, 772, 30

\bibitem[Dong et al.(2016)]{dong_2016} Dong, S., Shappee, B.~J., Prieto, J.~L., et al.\ 2016, Science, 351, 257

 \bibitem[Drago \& Pagliara(2015a)]{drago_2015a}    Drago, A., \& Pagliara, G.\ 2015a, \prc, 92, 045801
 
  \bibitem[Drago \& Pagliara(2015b)]{drago_2015b}    Drago, A., \& Pagliara, G.\ 2015b, arXiv:1509.02134
    
 \bibitem[Furusawa et al.(2015a)]{furusawa_2015a}       Furusawa, S., Sanada,  T., \& Yamada, S.\ 2015a, arXiv:1511.08148

 \bibitem[Furusawa et al.(2015b)]{furusawa_2015b}        Furusawa, S., Sanada,  T., \& Yamada, S.\ 2015b, arXiv:1511.08153
    
\bibitem[Gal-Yam(2012)]{galyam_2012} Gal-Yam, A. 2012, Sci, 337, 927

\bibitem[Gilkis et al.(2016)]{gilkis_2016}    Gilkis, A., Soker, N., \& Papish, O.\ 2016, \apj, 826, 178

\bibitem[Herzog \& R\"opke(2011)]{herzog_2013} Herzog, M., \& R\"opke, F. K. 2011, Phys. Rev. D, 84, 083002

\bibitem[Horvath \& Benvenuto(1988)]{horvath_1988} Horvath,  J. E. \& Benvenuto, O. G. 1988, Phys. Lett. B, 213, 516

\bibitem[Inserra et al.(2013)]{inserra_2013} Inserra, C., Smartt,  S.~J., Jerkstrand, A., et al.\ 2013, \apj, 770, 128

\bibitem[Iwazaki(2005)]{iwazaki_2005} Iwazaki, A.\ 2005, \prd, 72, 114003 

\bibitem[Jaikumar et al.(2007)]{jaikumar_2007} Jaikumar, P., Meyer, B.~S., Otsuki, K., \& Ouyed, R.\ 2007, \aap, 471, 227

\bibitem[Kasen \& Bildsten(2010)]{kasen_2010}   Kasen, D., \& Bildsten, L. 2010, \apj, 717, 245

\bibitem[Ker{\"a}nen et al.(2005)]{keranen_2005}  Ker{\"a}nen, P.,  Ouyed, R., \& Jaikumar, P.\ 2005, \apj, 618, 485

  \bibitem[Kostka et al.(2014a)]{kostka_2014a}  Kostka, M., Koning, N., Leahy, D., Ouyed, R., \& Steffen, W.\ 2014a, Revista Mexicana de Astronom\'ia y Astrof\'isica, 50, 167
  
  \bibitem[Kostka et al.(2014b)]{kostka_2014b}      Kostka, M., Koning, N., Shand, Z., Ouyed, R., \& Jaikumar, P.\ 2014b, \aap, 568, A97
  
 \bibitem[Kozyreva \& Blinnikov(2015)]{kozyreva_2015}    Kozyreva, A., \& Blinnikov, S. 2015, MNRAS, 454, 4357, 1510.00439

\bibitem[Leahy \& Ouyed(2008)]{leahy_2008}   Leahy, D., \& Ouyed, R.\ 2008, \mnras, 387, 1193

\bibitem[Leloudas et al.(2016)]{leloudas_2016} Leloudas, G., Fraser, M., Stone, N.~C., et al.\ 2016, arXiv:1609.02927

\bibitem[Margutti et al.(2016)]{margutti_2016}  Margutti, R., Metzger, B.~D., Chornock, R., et al.\ 2016, arXiv:1610.01632

\bibitem[Nicholl et al.(2014)]{nicholl_2014}  Nicholl, M., Smartt, S.~J., Jerkstrand, A., et al.\ 2014, \mnras, 444, 2096

\bibitem[Moriya et al.(2015)]{moriya_2015}  Moriya, T.~J., Liu, Z.-W., Mackey, J., Chen, T.-W., \& Langer, N.\ 2015, \aap, 584, L5 

\bibitem[Niebergal et al.(2010)]{niebergal_2010}  Niebergal, B., Ouyed, R., \& Jaikumar, P. 2010, Phys. Rev. C, 82, 062801

\bibitem[Ouyed et al.(2002)]{ouyed_2002}   Ouyed, R., Dey, J., \& Dey, M. 2002, A\&A, 390, L39

\bibitem[Ouyed et al.(2005)]{ouyed_2005}    Ouyed, R., Rapp, R.,  \& Vogt, C.\ 2005, \apj, 632, 1001

\bibitem[Ouyed \& Leahy(2009)]{ouyedleahy_2009}  Ouyed, R., \& Leahy, D.\ 2009, ApJ, 696, 562

\bibitem[Ouyed et al.(2009)]{ouyed_2009}  Ouyed, R., Niebergal, B., 
\& Jaikumar, P.\ 2009,  ``Predictions for signatures of the quark-nova in superluminous supernovae"  in  Proceedings of the Compact Stars in the QCD Phase Diagram II, http://vega.bac.pku.edu.cn/rxxu/csqcd.htm\ [arXiv:0911.5424]

 \bibitem[Ouyed et al.(2012)]{ouyed_2012}     Ouyed, R., Kostka, M., Koning, N., Leahy, D.~A., \& Steffen, W.\ 2012, \mnras, 423, 1652

\bibitem[Ouyed, Niebergal \& Jaikumar(2013)]{ouyed_2013a}  Ouyed, R., Niebergal, B., 
\& Jaikumar, P.\ 2013,  ``Explosive Combustion of a NS into a QS: the non-premixed scenario"  in  Proceedings of the Compact Stars in the QCD Phase Diagram III, http://www.slac.stanford.edu/econf/C121212/\ [arXiv:1304.8048]

\bibitem[Ouyed, Koning \& Leahy(2013)]{ouyed_2013b} Ouyed, R., Koning, N., \& Leahy, D.\ 2013, Research in Astronomy and Astrophysics, 13, 1463-1470

\bibitem[Ouyed \& Leahy(2013)]{ouyed_leah_2013} Ouyed, R., \& Leahy, D.\ 2013, Research in Astronomy and Astrophysics, 13, 1202-1206

\bibitem[Ouyed et al.(2015a)]{ouyed_2015a}  Ouyed, R., Leahy, D., \& Koning, N.\ 2015a, \mnras, 454, 2353 

\bibitem[Ouyed et al.(2015b)]{ouyed_2015b} Ouyed, R., Leahy, D.,  \& Koning, N.\ 2015b, \apj, 809, 142 

\bibitem[Ouyed et al.(2016)]{ouyed_2016} Ouyed, R., Leahy, D.,  \& Koning, N.\ 2016, \apj, 818, 77

 \bibitem[Pagliara et al.(2013)]{pagliara_2013}     Pagliara, G., Herzog,  M., {\ R\"o}pke, F.~K.\ 2013, \prd, 87, 103007

\bibitem[Piro(2015)]{piro_2015}  Piro, A.~L.\ 2015, \apjl, 808,  L51

\bibitem[Pastorello et al.(2010)]{pastorello_2010} Pastorello, A., Smartt, S. J., Botticella, M. T., et al. 2010, \apj, 724, L16

\bibitem[Quimby et al.(2011)]{quimby_2011} Quimby, R. M., Kulkarni, S. R., Kasliwal, M. M., et al. 2011, \nat, 474, 487

\bibitem[Sorokina et al.(2015)]{sorokina_2015} Sorokina, E.,  Blinnikov, S., Nomoto, K., Quimby, R., 
\& Tolstov, A.\ 2015, arXiv:1510.00834

\bibitem[Vogt et al.(2004)]{vogt_2004} Vogt, C., Rapp, R.,  \& Ouyed, R.\  2004, Nuclear Physics A, 735, 543

\bibitem[Wang et al.(2015)]{wang_2015} Wang, S.~Q., Liu, L.~D.,  Dai, Z.~G., Wang, L.~J., \& Wu, X.~F.\ 2015, arXiv:1509.05543

\bibitem[Weber(2005)]{weber_2005}  Weber, F. 2005,  Progress in Particle and Nuclear Physics,  54, 193

\bibitem[Witten(1984)]{witten_1984}  Witten, E. \ 1984,  Phys. Rev. D, 30, 272 

\bibitem[Woosley(2010)]{woosley_2010} Woosley, S. 2010, \apjl, 719, L204



\end{thebibliography}
\end{document}